\documentclass[twoside,aps,prb,twocolumn]{revtex4}
\pdfoutput=1
\usepackage{times}
\usepackage[T1]{fontenc}
\usepackage[latin1]{inputenc}
\usepackage{amsmath}
\usepackage{graphicx}
\usepackage{amssymb}

\newcommand{\citeasnoun}[1]{Ref.~\onlinecite{#1}}

\newcommand{\figref}[1]{Fig.~\ref{fig:#1}}

\newcommand{\<}{\langle}
\renewcommand{\>}{\rangle}
\newcommand{\p}{\partial}

\newcommand{\R}{\mathbb{R}}

\newcommand{\be}{\begin{equation}}
\newcommand{\ee}{\end{equation}}

\renewcommand{\vec}[1]{\mathbf{#1}}

\begin{document}

\preprint{Preprint}

\title{Sufficient conditions for two-dimensional localization by \\
arbitrarily weak defects in periodic potentials with band gaps}

\author{Arthur Parzygnat}

\affiliation{Department of Physics, Queens College, Flushing NY 11367}

\author{Karen K. Y. Lee}

\affiliation{Department of Electrical Engineering and Computer Science, Massachusetts Institute of Technology, Cambridge MA 02139}

\author{Yehuda Avniel}

\affiliation{Research Laboratory of Electronics, Massachusetts Institute of Technology, Cambridge MA 02139}

\author{Steven G. Johnson}

\email{stevenj@math.mit.edu}

\affiliation{Department of Mathematics, Massachusetts Institute of Technology, Cambridge MA 02139}

\begin{abstract}
\label{sec:abstract}
We prove, via an elementary variational method, 1d and 2d localization
within the band gaps of a periodic Schr\"odinger operator for any
mostly negative or mostly positive defect potential, $V,$ whose depth
is not too great compared to the size of the gap. In a similar way, we
also prove sufficient conditions for 1d and 2d localization below the
ground state of such an operator. Furthermore, we extend our results
to 1d and 2d localization in $d$ dimensions; for example, by a linear or
planar defect in a 3d crystal. For the case of $D$-fold degenerate
band edges, we also give sufficient conditions for localization of up
to $D$ states.
\end{abstract}

\maketitle

\markboth{Published in {\it Phys Rev. B}, vol. 81, p. 155324 (2010).}{Published in {\it Phys Rev. B}, vol. 81, p. 155324 (2010).}

\section{Introduction}
\label{sec:introduction}

Localization by impurities in periodic potentials with spectral gaps
(band gaps) is a central topic in solid-state physics and
semiconductor devices,\cite{AM} and it has direct analogues for other
propagating-wave systems, such as for photonic crystals in
electromagnetism.\cite{JoannopoulosJo08-book} We prove that, in one
and two dimensions (1d and 2d), localized solutions must arise in the
gaps of a periodic Schr\"odinger operator for any ``mostly'' negative
or mostly positive defect potential whose depth is not too great
compared to the size of the gap. To our knowledge, this is the first
rigorous sufficient condition of this sort in 2d, aside from informal
arguments based on effective-mass free-electron models close to
quadratic gap edges,\cite{MG} extending an earlier theorem for 1d
localization in gaps,\cite{Prodan06} and is quite different from the
many rigorous asymptotic gap-localization results in the limit of very
\emph{strong} defect potentials.\cite{DH,GeSi88,He,Sa,HuSi,Po} In
addition to localization in gaps, we also prove 1d and 2d localization
below the ground state of a periodic potential for any mostly negative
defect potential, extending earlier known results for localization in
vacuum for any mostly negative potential\cite{Si,Ec,YL}, localization
in the periodic case but for a strictly non-positive defect
potential,\cite{HuR,FSW} and 1d localization for mostly negative
defects.\cite{Prodan06} Furthermore, we extend our results to 1d and
2d localization in $d$ dimensions; for example, establishing
localization for a linear or planar defect in a 3d crystal. For the
case of $D$-fold degenerate band edges, we show localization of $D$
bound states for definite-sign defect potentials, and more generally
relate the number of localized modes to the signs of the eigenvalues
of a $D \times D$ matrix.  Our proofs rely only on elementary
variational eigenvalue bounds, generalizing an approach developed in
\citeasnoun{YL}.

One-dimensional localization in vacuum by an arbitrary attractive
potential is easy to prove by a variational method, at the level of an
undergraduate homework problem,\cite{LL} while a minimum non-zero
depth for a potential well is required for localization in vacuum in
three dimensions.\cite{BD} The 2d case is more
challenging to analyze. The first proof of 2d
localization in vacuum by an arbitrary attractive potential was
presented in 1976 by Simon,\cite{Si} using more sophisticated
techniques that also lead to asymptotic properties of the bound
states. An elementary variational proof in vacuum was proposed by
Picq\cite{Picq82,BB} (and was adapted to Maxwell's equations in
optical fibers by Bamberger and Bonnet\cite{BB}), while a different
variational proof was independently developed by Yang and
de~Llano.\cite{YL} An informal asymptotic argument utilizing
properties of the vacuum Green's function was presented by
Economou.\cite{Ec} On the other hand, in the case of periodic
potentials, rigorous results for the existence of bound states from
weak defect potentials $V$ are more limited. Frank
et~al.\cite{FSW}\ analyzed the case of potentials $V \le 0$ localizing
at energies below the ground state of a 2d periodic Schr\"odinger
operator; they not only proved that a localized state exists, but also
bounded its energy in terms of a related vacuum localization problem.
Here, we use a different technique, based on the variational method of
\citeasnoun{YL}, to prove existence of localized modes in 1d and 2d
for any indefinite-sign but ``mostly'' negative defect
potential~$V$. This is closely related to our generalization of
\citeasnoun{YL} to index guiding in the periodic Maxwell's
equations.\cite{KAJ} (Indefinite-sign localization was also considered
for \emph{discrete} Schr{\"{o}}dinger operators.\cite{DHu}) As for
localization in arbitrary gaps, however, we are not aware of published
rigorous results in 2d that are valid for weak $V$.
Prodan\cite{Prodan06} proved that arbitrarily weak defects localize
states in 1d gaps, using an asymptotic Birman--Schwinger technique
similar to \citeasnoun{Si}, imposing a mostly negative/positive
condition on the defect potential identical to the one we use below;
Prodan's result also applies to localization below the ground state of
a periodic 1d potential.  Various authors have shown that for
\emph{strong} defect potentials, those of the form $\lambda V$ where
$\lambda \to \infty,$ there exists a bound on the number of
eigenvalues crossing the gap.\cite{DH,GeSi88,He,HuSi,Sa} Localization
has also been proved in the limit of high-order gaps in
1d.\cite{GeSi93} Another common, albeit somewhat informal, approach to
gap localization is to consider localization for energies close to a
non-degenerate quadratic band edge, making an effective-mass
approximation and then quoting the results for vacuum.\cite{MG} Our
proof of localization in gaps is non-asymptotic, does not assume a
particular form of the band edge, and is an extension of the
elementary variational technique of \citeasnoun{YL} for localization
in vacuum. The trick behind the variational proof is to take the
mid-gap energy, $E_g$, of the perturbed Schr\"odinger operator, $H$,
and transform the question of energies near $E_g$ into an extremal
eigenvalue problem. There are two typical ways to do this. One is to
consider $(H-E_g)^{-1},$ which seems closely related to the Green's
function method of \citeasnoun{Ec} and to the Birmann--Schwinger
condition,\cite{HeR} but such an operator is hard to evaluate
explicitly and mainly lends itself to asymptotic analyses. Another
method is to consider $(H-E_g)^2$, recently used for another
variational-type localization proof by \citeasnoun{KO}, and it is this
method that we adopt here. The same techniques are used in numerical
methods to iteratively compute eigenvalues in the interior of the
spectrum of a large matrix, where $(H-E_g)^{-1}$ corresponds to
well-known shift-and-invert methods,\cite{Bai00} and where $(H-E_g)^2$
has also been used (but is computationally suboptimal because it
squares the condition number).\cite{JohnsonJo01,WangZu94} Other
possible techniques\cite{GeSi93,HuSi,We} have been suggested to
us\cite{Pushnitski-unpub} for proving such a theorem, but we are not
aware of any published results for this problem other than the 1d
result of \citeasnoun{Prodan06}.  Localization by weak defects is also
related to self-focusing of solitons by nonlinearities, which was
recently considered for spectral gaps in periodic
potentials.\cite{IlanWe10}

The contents of the individual sections are summarized as follows.
Sec.~\ref{sec:bsbi} provides a simple variational proof of the fact
that any arbitrary mostly negative and sufficiently localized defect
induces at least one bound state below the infimum of the spectrum of
a 2d periodic Schr\"odinger operator. The 1d case is proved in a
similar fashion.  Sec.~\ref{sec:dDim} gives a generalization of that
result by allowing the unperturbed Hamiltonian to be periodic in $d$
dimensions and our defect potential to localize in two (or one)
dimensions, but be periodic (with the same periodicity as the
unperturbed potential) in the other $d-2$ (or $d-1)$ dimensions.
Sec.~\ref{sec:bssg} presents our main results: it gives a sufficient
condition for the existence of a bound state in a spectral gap of a 2d
periodic Schr\"odinger operator. The results from Sec.~\ref{sec:dDim}
can be applied to this case as well to allow periodicity in any
arbitrary extra number of dimensions (although the solution is only
localized in two of these dimensions). The case of bound states in a
gap confined along one dimension is proved in a similar way.
Sec.~\ref{sec:remarks} describes the case of degenerate band edges
as well as some other possible generalizations.
Sec.~\ref{sec:future} closes with some concluding remarks and
possible future directions.

\section{Bound States Below the Ground State of a Periodic Potential}
\label{sec:bsbi}

Notation: In this section, unless otherwise stated, the symbol $\int f$ will stand for the integral over all of $\R^2$ of the real-valued function $f.$ 

The proof of the following will be a little simpler than the one in
Sec.~\ref{sec:bssg}. However, the ideas used here are almost the
same as those of the proof in Sec.~\ref{sec:bssg}, so it is hoped
that after going through this proof, the reader will easily follow the
latter. Note that the theorem in this section has also been proved by other
methods, but only for defect potentials that are strictly nonpositive.\cite{HuR,FSW} This theorem will be slightly generalized to
allow for a defect potential that is localized in two dimensions, but
has an arbitrary periodicity in all other dimensions, in Sec.~\ref{sec:dDim}.

\subsection{Problem statement}
\label{sub:bsbithm}
Suppose we start with an unperturbed Hamiltonian,
\begin{equation}
H_0 = - \nabla^2 + V_0,
\label{eq:H0}
\end{equation}
where $V_0$ is a periodic potential (possible generalization to
non-periodic potentials is discussed in Sec.~\ref{sub:periodicV0}),
which has a minimum energy eigenvalue $E_0$ with at least one
(degeneracy will be explored in Sec.~\ref{sub:degeneracy})
corresponding ``generalized'' eigenfunction $\psi_0$ of the Bloch
form:\cite{AM} a periodic function with the same periodicity as $V_0$
multiplied by $e^{i \vec{k} \cdot \vec{x}},$ giving a bounded
$|\psi_0|.$ We introduce a localized, indefinite (varying sign)
defect, $V,$ giving a new Hamiltonian $H = H_0 + V,$ satisfying the
two conditions
\begin{equation}
\int V |\psi_0|^2 < 0 \quad \text{and}
\label{eq:bsbiV1}
\end{equation}
\begin{equation}
\lim_{r \to \infty} V(r, \theta) = 0.
\label{eq:bsbiV2}
\end{equation}
Such a $V_0$ and $V$ are shown schematically in \figref{FigI} along
with a typical spectrum showing $E_0$ (and also possibly a gap, which
is considered in Sec. {\ref{sec:bssg}). We prove that conditions
  (\ref{eq:bsbiV1}) and (\ref{eq:bsbiV2}) are sufficient to guarantee
  the existence of a bound state with energy lower than $E_0$.
  [Strictly speaking, in order for an energy $< E_0$ to be a ``bound''
    state, the potential $V$ must be sufficiently localized so as to
    not alter the essential spectrum (energies of non-localized modes)
    of $H_0$; (\ref{eq:bsbiV2}) does not appear to be strong
    enough. Weyl's theorem provides a sufficient condition of
    relatively compact $V$, such as the Kato class of mostly
    square-integrable potentials, which seems to include most
    realistic localized defects.\cite{HislopSi96,Sc}] The proof is
  essentially a generalization of that in \citeasnoun{YL}.
\begin{figure}
\center
\includegraphics[width=1\columnwidth]{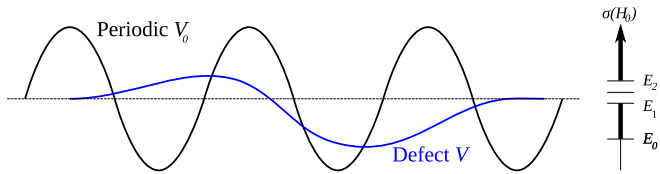}
\caption{(Color online) Example of a periodic potential $V_0$ and defect potential $V$ with the spectrum $\sigma(H_0)$ located at the right. The spectrum of the unperturbed Schr{\"{o}}dinger operator has an infimum $E_0$, and may also have a gap in the set $(E_1,E_2)$. \label{fig:FigI}}
\end{figure}

\subsection{Proof of 2d localization}
\label{sub:bsbipf}
Recall that the variational principle (or min--max principle) states
that the Rayleigh quotient $\< \psi, H \psi \> / \< \psi, \psi \>$,
where $\< \psi, \phi \> \equiv \int \psi^* \phi$ denotes the inner
product between $\psi$ and $\phi$ (and $\Vert \psi \Vert^2$ denotes
$\< \psi, \psi \>$), for any trial function $\psi$ (in the appropriate
Sobolev space) is an upper bound for the ground-state eigenvalue of a
Hermitian operator $H$.\cite{Gr,Sh,RSIV} Therefore, to prove the
existence of an eigenvalue less than $E_0$ (and thus a bound state),
it suffices to find a $\psi$ such that
\begin{equation}
E[\psi] \equiv \cfrac{ \< \psi, (H - E_0) \psi \> }{ \< \psi, \psi \> } < 0.
\end{equation}
The key point is to find a trial function
that will work even for an arbitrarily small defect $V.$ Motivated by
\citeasnoun{YL}, we use the following trial function parametrized by a positive
number $\alpha$:
\begin{equation}
\psi = \psi_0 \gamma, \quad \text{ where } \quad \gamma = e^{- (1+r)^{\alpha}}.
\end{equation}
Once the appropriate trial function is selected, the remaining
analysis is straightforward in principle---one simply plugs the trial
function into $E[\psi]$ and finds an $\alpha > 0$ where it is
negative.  The easiest way to do this is to take the $\alpha \to 0$
limit of the numerator: if this limit is negative, then there must
also exist a small $\alpha > 0$ where it is negative.  This process, which
requires some care in taking the limits (limits and integration cannot
be interchanged in general), is carried out as follows.

Note that $\gamma$ is already in polar coordinates and that $\psi$ is normalizable for all such $\alpha$ since $\psi_0$ is bounded. This trial function has the key physically motivated property that the limit of no localization, i.e. $\alpha \to 0,$ gives the unperturbed ground state $\psi_0.$ We write down the first two derivatives of $\gamma$ for future reference:
\begin{equation}
\gamma' \equiv \frac{\p \gamma}{\p r} = \hat{r} \cdot \nabla \gamma = - \alpha (1+r)^{\alpha-1} \gamma;
\label{eq:gamma'}
\end{equation}
\begin{equation}
\gamma'' \equiv \frac{\p^2 \gamma}{\p r^2} =  \alpha (1+r)^{\alpha - 2} \big [ \alpha (1+r)^{\alpha} - \alpha + 1 \big ] \gamma.
\label{eq:gamma''}
\end{equation}
When $H-E_0$ acts only on $\psi_0$ in $E[\psi],$ the result is zero. The remaining terms in the Rayleigh-quotient numerator, denoted by $U[\psi],$ come from $V$ and derivatives of $\gamma.$ After some algebraic simplifications, $U[\psi]$ is given by (see the Appendix) 
\begin{equation}\begin{split}
U[\psi] &= \< \psi, (H-E_0) \psi \> \\
&= \int V |\psi|^2 + \int |\psi_0|^2 \left [ \frac{1}{2} \nabla^2 (\gamma^2) - \gamma \nabla^2 \gamma \right ].
\label{eq:bsbiU1}
\end{split}\end{equation}
Using equations (\ref{eq:gamma'}) and (\ref{eq:gamma''}), we obtain
\begin{equation}
\nabla^2 \gamma 
	= \alpha (1+r)^{\alpha-2} \left [ \alpha (1+r)^{\alpha} - \alpha - \frac{1}{r} \right ] \gamma;
	\label{eq:Deltagamma}
\end{equation}
\begin{equation}
\nabla^2 (\gamma^2) = 2 \alpha (1+r)^{\alpha-2} \left [ 2 \alpha (1+r)^{\alpha} - \alpha - \frac{1}{r} \right ] \gamma^2.
\label{eq:Deltagamma2}
\end{equation}
Plugging these two formulas into $U[\psi]$ results in the concise form:
\begin{equation}
U[\psi] = \int V |\psi|^2 + \int |\psi_0|^2 \alpha^2 (1+r)^{2\alpha-2} \gamma^2.
\label{eq:bsbiU2}
\end{equation}
Note that the denominator of $E[\psi]$ (which is $\lVert \psi \rVert^2)$ is always positive and so does not affect the sign of $U[\psi].$ We want to show that $U[\psi],$ and thus $E[\psi],$ will be negative for some choice of $\alpha.$ This will be done by showing that the term on the right of equation (\ref{eq:bsbiU2}) tends to zero as $\alpha \to 0,$ while $\int V |\psi|^2$ will be negative in this limit.  Because $|\psi_0|^2$ is bounded, we have
\begin{multline}
\label{eq:maxpsi} 
\int |\psi_0|^2 \alpha^2 (1+r)^{2\alpha-2} \gamma^2 \\
        \le 2\pi \max\{ |\psi_0|^2 \}  \int_0^{\infty} \alpha^2 (1+r)^{2\alpha-2} \gamma^2 r \, dr \\
	=2\pi \max\{ |\psi_0|^2 \} \int_1^{\infty} \alpha^2 u^{2\alpha -2} \gamma^2 (u-1) du  \\
	\le 2\pi \max\{ |\psi_0|^2 \} \int_1^{\infty} \alpha^2 u^{2\alpha -1} \gamma^2 du, 
\end{multline}
where we have made the substitution $u = 1+r$ and then bounded the integral again. Hence, it suffices to show that the latter integral tends to zero. We calculate this integral explicitly via integration by parts:
\begin{multline}
\int_1^{\infty} \alpha^2 u^{2\alpha - 1} \gamma^2 du \\
= \left. -\frac{\alpha}{2} u^{\alpha} e^{- 2u^{\alpha}} \right|_{1}^{\infty} + \int_1^{\infty} \frac{\alpha^2}{2} u^{\alpha-1} e^{-2u^{\alpha}} du = \frac{3 \alpha}{4e^2}.
\label{eq:nice2}
\end{multline}
Taking the limit as $\alpha \to 0$ yields zero as claimed above. 

The leftover $\int V |\psi|^2$ term can be split into two parts.
Let $V = V^+ - V^-,$ where $V^+$ and $V^-$ are the positive and negative parts of $V.$
Then we have
\begin{equation}
\int |\psi|^2 V = \int |\psi|^2 V^+ - \int |\psi|^2 V^- ,
\end{equation}
where each $|\psi|^2 V^{\pm}$ is a monotonically increasing function as $\alpha$ decreases. This allows us to use Lebesgue's monotone convergence theorem\cite{Ru} to interchange the limit with the integration, arriving at
\begin{equation}
\lim_{\alpha \to 0} \int |\psi|^2 V^{\pm} = \int \lim_{\alpha \to 0} |\psi|^2 V^{\pm} = e^{-2} \int |\psi_0|^2 V^{\pm}
\label{eq:bsbi2}
\end{equation}
for each part. 
Putting all the information together, we have 
\begin{multline}
\lim_{\alpha \to 0} \left (  \int V |\psi|^2 + \int |\psi_0|^2 \alpha^2 (1+r)^{2\alpha-2} \gamma^2 \right ) \\
= e^{-2} \int |\psi_0|^2 V < 0
\label{eq:final-inf-eq}
\end{multline}
by our main assumption (\ref{eq:bsbiV1}). Hence, the variational principle says that there exists an eigenvalue below $E_0$ for the system and so the theorem is proved.

\subsection{Proof of 1d localization}
The case for 1d localization can be proved in an analogous way, with simpler calculations, by using the trial function $\psi_0 e^{ - \alpha x^2}.$ 

A closely related result in 1d was presented by Prodan:\cite{Prodan06}
for any defect potential $V$ satisfying~(\ref{eq:bsbiV1}), and for any
energy $E < E_0$, Prodan showed that there was some scaling $\lambda
V$ with $\lambda > 0$ such that a bound state with energy $E$ exists.
Furthermore, the limit $E \to E_0$ was shown to correspond to $\lambda
\to 0$, so that an arbitrarily weak potential
satisfying~(\ref{eq:bsbiV1}) must localize a state.

\section{2d Localization in $d$ Dimensions}
\label{sec:dDim}

We would like to extend these results to an unperturbed Hamiltonian
$H_0 = - \nabla^2 + V_0$ that is periodic in $d$ dimensions and where
the defect potential $V$ is localized along one or two dimensions and
is periodic in the other dimensions. A physical example of 2d
localization would be a linear defect or ``quantum wire'' in a 3d
crystal, localizing a wavefunction to propagate along the line,
whereas an example of 1d localization in 3d would be a planar defect.

\subsection{Periodicity of $V_0$ and $V$}
\label{sec:notation}
In this case, $V_0$ is a periodic potential in $d$ dimensions, while
$V$ is periodic only in $d-2$ dimensions but localized in the other
two. It is convenient to separate the periodic and nonperiodic
coordinates of $V$ by writing $V$ as $V(r, \theta, \vec{z}),$ where
$\vec{z} \in \R^{d-2},$ which is periodic in $\vec{z}$ with some
lattice vectors. The first two coordinates are in polar form for
convenience in defining what we mean by ``localization,'' which occurs
in $r$ only. $V_0,$ on the other hand, is periodic in \emph{all}
dimensions, but for convenience, we also write it as $V_0 (r, \theta,
\vec{z}).$ A schematic 2d example of such a $V$ and $V_0,$ where $V$
is localized in only one dimension and periodic in one other, is shown
in \figref{FigP}.  Note that it is irrelevant in our proof whether $V_0$ is
periodic in any $(r,\theta)$ plane, which only occurs if $(r,\theta)$
corresponds to a lattice plane\cite{AM} of $V_0$ (such an orthogonal
supercell exists only under certain conditions on the lattice
vectors\cite{Bucksch73}).
\begin{figure}
\center
\begin{centering}\includegraphics[width=0.7\columnwidth]{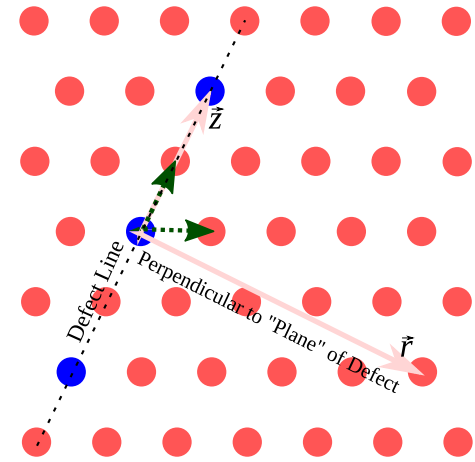}\par\end{centering}
\caption{(Color online) Linear defect for a periodic lattice (red circles) in two dimensions is created by adding a defect potential (blue circles) that is periodic along the $\vec{z}$ direction, commensurate with the periodicity of the underlying lattice.  This will localize a state with respect to the perpendicular directions, denoted $\vec{r}$.\label{fig:FigP}}
\end{figure}

The corresponding Laplacian for these coordinates is:
\begin{equation}
\nabla^2 = \nabla^2_{r,\theta} + \nabla^2_{\vec{z}} = \left (\frac{\p^2}{\p r^2} + \frac{1}{r} \frac{\p}{\p r} + \frac{1}{r^2} \frac{\p^2}{\p \theta^2} \right ) + \sum_{j=3}^d \frac{\p^2}{\p x_j^2}.
\end{equation}

In this section and the next, the symbol $\int f$ will be interpreted as
\begin{equation}
\int f \equiv  \int_0^{\infty} r \, dr \int_0^{2\pi} d\theta \int_{\Omega} d^{d-2} \vec{z} f(r, \theta, \vec{z}),
\end{equation}
where $\Omega$ is the primitive cell in $\vec{z}$. 

\subsection{Generalized localization below ground state}
\label{sub:gen}
Due to the periodicity of the potentials, we apply Bloch's theorem to reduce this problem to a unit cell in $\vec{z},$ in which case a ``localized'' mode is a point eigenvalue as in the previous section, and we can prove localization by an identical variational approach. 

In particular, all eigensolutions can be chosen in the Bloch form
$\psi_0 = \psi_0^k (r, \theta, \vec{z}) e^{i \vec{k} \cdot \vec{z}},$
where $\psi_0^k$ is periodic in $\vec{z}$ with the same periodicity as
$V$ (and $V_0$) and $\vec{k}$ is some Bloch
wavevector.\cite{AM} Substituting this into $H_0 \psi = E \psi,$ one
obtains the ``reduced'' Hamiltonian
\begin{equation}
H^k_0 = - \nabla^2 - 2 i \vec{k} \cdot \nabla_{\vec{z}} + \vec{k}^2 + V_0,
\label{eq:H0k}
\end{equation}
whose eigenvalues $E(\vec{k})$ are the $\vec{k}$-dependent band
structures. The domain is now only the primitive cell in $\vec{z}$
with periodic boundary conditions.  For each $\vec{k},$ there is a
minimum energy, $E_0 (\vec{k}),$ associated with each reduced
Hamiltonian $H^k_0,$ i.e. $H^k_0 \psi^k_0 = E_0 (\vec{k}) \psi^k_0$
for some Bloch wavefunction $\psi_0^k.$ ($\psi_0^k
e^{i\vec{k}\cdot\vec{z}}$ is merely one of the Bloch wavefunctions for
the underlying potential $V_0$ such that the projection of the original
$d$-dimensional Bloch wavevector onto the $\vec{z}$ dimensions gives
$\vec{k}$.  Therefore, $|\psi^k_0|$ is bounded.)  The claim we wish to
prove is that the above conditions (just as in Sec.~\ref{sec:bsbi})
guarantee the existence of a bound state with energy below
$E_0(\vec{k})$ for each $\vec{k}.$ Moreover, the previous theorem of
Sec.~\ref{sec:bsbi} becomes merely the special case when $\vec{k} =
0$ and there are no extra dimensions of periodicity.

\subsection{Proof of 2d localization in $d$ dimensions}
The proof follows the same outline as in Sec.~\ref{sec:bsbi} (just replace $\psi_0$ with $\psi^k_0).$ There are only two small differences. $H_0^k$ in (\ref{eq:H0k}) has the additional terms $2 i \vec{k} \cdot \nabla_{\vec{z}}$ and $\vec{k}^2,$ but $\nabla_{\vec{z}} \gamma = 0$ so this additional term only acts on $\psi_0^k,$ and is part of the $(H_0 - E_0) \psi_0^k = 0$ cancellation as before in (\ref{eq:bsbiU1}). The second difference is that these integrals are over $\R^2 \times \R^{d-2}$ (where the part over $\R^{d-2}$ is actually just over a primitive cell), instead of over $\R^2.$ This simply means that, in (\ref{eq:maxpsi}), instead of factoring out $\max \{ |\psi_0^k|^2 \}$ in the bound, we factor out $\max \{ |\psi_0^k|^2 \}$ multiplied by the volume of the primitive cell (since the remaining integrand is independent of $\vec{z}).$

\subsection{Proof of 1d localization}
The analysis is the same except we now have $V_0 = V_0(x, \vec{z})$ and $\psi = \psi_0^k e^{i \vec{k} \cdot \vec{z}} e^{- \alpha x^2}.$ 

\section{Bound States Within a Band Gap}
\label{sec:bssg}

Now, we suppose that there is a gap in the spectrum of $H_0^k$ (at a particular $\vec{k}$ in $d$ dimensions) from $E_1$ up to $E_2$ (with corresponding band edge Bloch wavefunctions $\psi_1^k$ and $\psi_2^k)$ as shown in \figref{FigI}, and we prove sufficient conditions for $V$ to localize a state along one or two dimensions with an energy in the gap (with periodicity as defined in the previous section). Intuitively, if $V$ is mostly negative, then a state will be pulled \emph{down} into the gap from the higher energy edge $E_2.$ If it is mostly positive instead, then a state will be pushed \emph{up} into the gap from the lower energy edge $E_1.$ On the other hand, if the potential is too strong, then the band edge state will be pulled/pushed all the way across the gap and will cease to be localized, hinting that we may need some upper bound on the strength of the defect potential even if there is no lower bound. This will be made quantitative in Sec.~\ref{sub:thm}. 

The idea behind the actual proof is to translate it to a simpler problem so that the methods of Sec.~\ref{sec:bsbi} can be used. That is, we must somehow alter the operator $H$ so that we are localizing below the ``ground state'' of an altered $H_0$ and can use the variational principle. This is achieved by  considering $(H-E_g)^2,$ where $E_g = (E_2-E_1)/2,$ instead of just $H,$ which transforms the spectrum as shown in \figref{FigG}. This idea comes from the localization proofs of \citeasnoun{KO} as well as from numerical techniques.\cite{JohnsonJo01,WangZu94} The trial function is motivated once again by \citeasnoun{YL}.  

\begin{figure}
\center
\includegraphics[width=1\columnwidth]{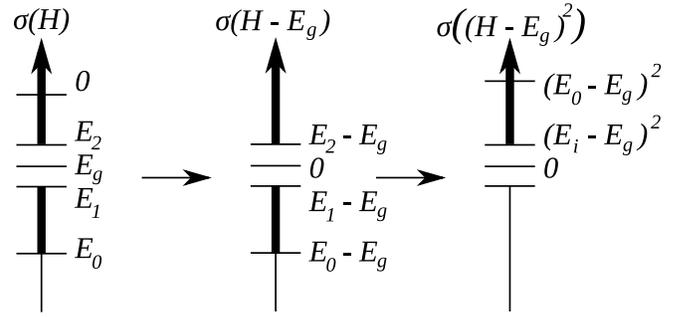}
\caption{The leftmost plot is a schematic picture of the spectrum $\sigma(H)$ (with some arbitrary origin $0$). The first step is to shift the center of the gap to the zero position by subtracting $E_g.$ The middle plot shows $\sigma(H-E_g).$ Then, the operator is squared and all energies become positive. The final plot shows $\sigma[(H-E_g)^2]$: any eigenvalue introduced into the gap of the original operator will now be an extremal eigenvalue of the shifted and squared operator. \label{fig:FigG}}
\end{figure}

\subsection{Statement of the theorem}
\label{sub:thm}
For ease of notation, we will omit the $k$ superscripts in this section. 

We will prove gap localization under the following conditions. The intuition that $V$ should be mostly negative or positive corresponds to our condition:
\begin{equation}
\int V |\psi_1|^2 < 0 \quad  \text{ and/or } \quad \int V |\psi_2|^2 > 0. \label{eq:Vprop1}
\end{equation}
The intuition that $V$ cannot be too big corresponds to:
\begin{equation}
2 | E_i - E_g | \left | \int V |\psi_i|^2 \right | > \int V^2 |\psi_i|^2, \label{eq:Vprop3}
\end{equation}
where $i$ depends on which case of (\ref{eq:Vprop1}) was satisfied; that
is, from which band edge we are pulling a localized state.
[Conditions (\ref{eq:Vprop1})--(\ref{eq:Vprop3}) can be merged into a
  single condition, below, that (\ref{eq:Vtermsrem}) is negative.] We
also require that $V$ be sufficiently localized in $r$, corresponding
to the condition:
\begin{equation}
\int |V| |\psi_i|^2 < \infty \quad  \text{ and } \quad \int V^2 |\psi_i|^2 <\infty. \label{eq:Vprop2}
\end{equation}
These conditions are sufficient to guarantee a bound state \emph{in the gap} for the perturbed Hamiltonian $H = H_0 + V,$ i.e. a localized state in the $(r, \theta)$ dimensions.

\subsection{Proof of 2d localization}
\label{sub:pf}
Considering one band edge at a time, we will prove localization of a bound state ``pulled'' from the band edge $E_i$ for $i=1,2.$ 
The proof will be split up into several steps to make it easier to follow. The method employed is the same as that in Sec.~\ref{sec:bsbi}.

The variational principle in this problem can be used after shifting the center of the gap to the zero position and squaring the resulting operator thus making it sufficient to find a normalizable trial function $\psi$ such that
\begin{equation}
\dfrac{ \< \psi,  (H-E_g)^2 \psi \> }{ \< \psi, \psi \> } < (E_i - E_g)^2 \equiv (\Delta E)^2 \label{eq:Goal}
\end{equation}
or equivalently,
\begin{equation}
\lVert (H - E_g) \psi \rVert^2 - (\Delta E)^2 \lVert \psi \rVert^2 < 0. \label{eq:Cond}
\end{equation} 

Consider the trial function $\psi = \gamma \psi_i,$ where $\gamma = \exp \{-(1+r)^{\alpha} \}$ as before. Similar to Sec.~\ref{sec:bsbi}, we will show that this $\psi$ will satisfy condition (\ref{eq:Cond}) for some small $\alpha$ providing conditions (\ref{eq:Vprop1})--(\ref{eq:Vprop2})  are also satisfied.

After some algebraic manipulation and using the fact that $H_0 \psi_i = E_i \psi_i,$ the left-hand side of (\ref{eq:Cond}) becomes (as shown in more detail in the Appendix)
\begin{gather}
\lVert (H - E_g) \psi \rVert^2 - (\Delta E)^2 \lVert \psi \rVert^2 = \qquad \qquad \qquad \qquad\nonumber \\
(\Delta E) \int |\psi_i|^2 \left [ \nabla^2 (\gamma^2) - 2 \gamma \nabla^2 \gamma \right ]  \label{eq:Top}
\\
\begin{split}
\qquad \qquad &{} + 2(\Delta E) \int V |\psi|^2 + \lVert V \psi \rVert^2 \\
&{} - \int V \left [ \nabla |\psi_i|^2 \cdot \nabla (\gamma^2) - 2 |\psi_i|^2 \gamma \nabla^2 \gamma \right ]  \label{eq:Middle}
\end{split} \\
\begin{split}
&{} + 4 \lVert \nabla \psi_i \cdot \nabla \gamma \rVert^2 + \lVert \psi_i \nabla^2 \gamma \rVert^2 \; \; \; \; \;\\
&{} + 2 \int \nabla^2 \gamma \nabla |\psi_i|^2 \cdot \nabla \gamma. \label{eq:Bottom}
\end{split}
\end{gather}

It is a bit cumbersome, but we can separate the integrals into four main groups to avoid calculating each one individually.  

The first group consists of (\ref{eq:Top}), but we have already shown that this integral tends to zero as $\alpha \to 0$ in Sec.~\ref{sec:bsbi}---see equations (\ref{eq:maxpsi}) and (\ref{eq:nice2}).

The second group consists of all the integrals which involve $V.$ These are all contained in (\ref{eq:Middle}). Taking the limit as $\alpha \to 0$ is allowed to pass through the integrals because everything is bounded by either a constant times $V$ or a constant times $V^2$, and our assumption (\ref{eq:Vprop2}) allows us to use Lebesgue's dominated convergence theorem\cite{Ru} to commute the limits with integration.
The limit of these integrals as $\alpha \to 0$ is
\begin{equation}
e^{-2} \int V^2 |\psi_i|^2 + 2 (E_i - E_g) e^{-2} \int V |\psi_i|^2, 
\label{eq:Vtermsrem}
\end{equation}
where the rightmost integrand of (\ref{eq:Middle}) vanishes for $\alpha \to 0$ because differentiating $\gamma$ results in at least one $\alpha$ factor. Equation (\ref{eq:Vtermsrem}) is strictly negative under conditions (\ref{eq:Vprop1}) and (\ref{eq:Vprop3}). Namely, the $V^2$ term is smaller than the $V$ term by (\ref{eq:Vprop3}), and (\ref{eq:Vprop1}) implies that the $(E_i-E_g) V |\psi_i|^2$ integral is negative for either $i=1$ or $i=2.$ 

We now move on to the final terms: everything in (\ref{eq:Bottom}). We wish to show that they all tend to zero as $\alpha \to 0,$ and we can do this easily by concentrating on the term that decays the \emph{most slowly} with respect to $r,$ so that all the other terms which decay faster clearly go to zero as well (provided that the terms have the same or higher order for the $\alpha$ factor in front). The three integrals of (\ref{eq:Bottom}), dropping bounded terms such as $|\psi_i|^2,$ are [from left to right in~(\ref{eq:Bottom})]:
\begin{align}
(\nabla \gamma)^2 &= \alpha^2 (1+r)^{2\alpha - 2} \gamma^2 , \label{eq:Dg1} \\
(\nabla^2 \gamma)^2 &= \alpha^2 (1+r)^{2\alpha - 4} \left [ \alpha(1+r)^{\alpha} - \alpha - \frac{1}{r} \right ]^2 \gamma^2 , \label{eq:Dg2} \\
\nabla^2 \gamma \nabla \gamma &= \alpha^2 (1+r)^{2\alpha - 3} \left [ \alpha(1+r)^{\alpha} - \alpha - \frac{1}{r} \right ] \gamma^2. \label{eq:DgNg} 
\end{align}
Each has at least an $\alpha^2$ factor in front. Upon inspection, we find that the most slowly decaying term out of these three is the $(\nabla \gamma)^2$ term, which goes as $1/r^2$ in the limit of $\alpha \to 0$ and its limit is
\begin{equation}
\lim_{\alpha \to 0} \int_0^{\infty} \alpha^2 (1+r)^{2\alpha -2} \gamma^2 r \, dr \to 0,
\end{equation}
as was already shown in (\ref{eq:maxpsi}) and (\ref{eq:nice2}). Since (\ref{eq:Dg1}) dominates (\ref{eq:Bottom}), the other terms are all asymptotically bounded by some constant times this integral and hence must also have integral zero in the limit as $\alpha \to 0.$  

What has been shown is that every term from the left-hand side of (\ref{eq:Cond}) vanishes except for the one term (\ref{eq:Vtermsrem}), which is negative. This establishes the existence of a bound state.

\subsection{Proof of 1d localization}
The case for 1d localization can be proved in an analogous way, with simpler calculations, by using the trial function $\psi_0 e^{ - \alpha x^2}$ just as before.  

A closely related result in 1d was presented by Prodan:\cite{Prodan06}
for any defect potential $V$ satisfying~(\ref{eq:Vprop1}), and for any
energy $E$ in the gap, Prodan showed that there was some scaling
$\lambda V$ with $\lambda > 0$ such that a bound state with energy $E$
exists.  Furthermore, the limit as $E$ approaches the band edge
corresponding to~(\ref{eq:Vprop1}) was shown to correspond to $\lambda
\to 0$, so that the limit of an arbitrarily weak potential
satisfying~(\ref{eq:Vprop1}) must localize a state in the gap.

\section{Some further generalizations}
\label{sec:remarks}

\subsection{Necessity of periodicity?}
\label{sub:periodicV0}
In all our derivations, we did not actually explicitly use the fact that $V_0$ was periodic in the dimensions where localization would take place. All we used were a few of the properties of periodic potentials. These are:
\begin{itemize}
\item The energies are bounded from below.
\item There may be a finite gap inside the continuous spectrum.
\item The generalized eigenfunctions corresponding to the infimum/gap-edge energies are bounded (and their derivatives are bounded).
\end{itemize}

For gap localization, we also assumed that the squared operator was well-defined, making application of this theorem to  delta-function potentials (Kronig--Penney models) appear problematic, although it is possible that the difficulty is surmountable with a sufficiently careful limiting process. (In physical contexts, the difference between a theoretical infinite-depth $V_0$ and a finite-depth approximation seems scarcely relevant.)  Also, we assumed coinciding essential spectra for $H_0$ and $H$ in order to utilize the variational principle. This means that there are some restrictions on how large of a perturbation $V$ can be. However, Weyl's theorem states one sufficient condition,\cite{HislopSi96,Sc} which appears to be satisfied for most physically interesting $V$'s; see also the comments after~(\ref{eq:bsbiV2}).

The third assumption listed above may be more challenging to prove for
non-periodic potentials. We assumed that the energies at the infimum
of the spectrum and/or the edges of spectral gaps correspond to
eigenvalues with bounded generalized eigenfunctions ($\psi_0$ or
$\psi_i$). (This corresponds to the requirement of a regular ground
state in \citeasnoun{FSW}.) For periodic potentials, the existence of
a band-edge solution for $V_0$ follows from the well-known continuity
of the band diagram as a function of the Bloch wavevector
$\vec{k}$. For non-periodic potentials $V_0$, however, this is not
necessarily true. For example, for the 1d half-well potential $V_0(x)
= \infty$ for $x<0$ and $=0$ for $x>0$, the eigenfunctions
$\sin(\kappa x)$ do not have a nonzero infimum-energy solution for
$\kappa \to 0$, and correspondingly it is well known that any
perturbation $V$ must exceed some threshold depth before a bound state
appears in that case---this is mathematically equivalent to the
appearance of an odd bound-state solution $\psi(-x)=-\psi(x)$ for $V_0
= 0$ and an even perturbation $V(-x)=V(x)$, which requires a threshold
depth since the lowest-energy bound state in that case is even. It is
not clear to us under what conditions the requisite infimum/gap-edge
solutions exist for more general potentials, such as quasiperiodic
potentials $V_0$, although some examples are given in
\citeasnoun{FSW}.

\subsection{Degeneracy at the band edges}
\label{sub:degeneracy}
As mentioned earlier, it could happen that there are multiple (degenerate) linearly independent $\psi_i$'s corresponding to a given energy $E_i$ at an edge of the gap and/or at the infimum of the spectrum. Our proof in the preceding sections is unaffected---there must still be \emph{at least one} bound state localized by a suitable $V$, as long as the requisite conditions [(\ref{eq:bsbiV1}) or (\ref{eq:Vprop1}--\ref{eq:Vprop2})] hold for at least one of the degenerate $\psi_i$ wavefunctions. Intuitively, however, one might expect to be able to prove a stronger theorem in this case---if $E_i$ is $D$-fold degenerate, can one show that $D$ localized states are formed by a suitable $V$? 

To prove the existence of more than one localized state, we can employ a generalization of the min--max theorem. For a single localized state, our proof involved the fact that the ground-state eigenvalue of a Hermitian operator $O$ is bounded above by the Rayleigh quotient $Q[\psi] = \langle \psi, O \psi \rangle / \langle \psi, \psi \rangle$ for any $\psi$.  The generalization of this fact is that the $n$-th eigenvalue $\lambda_n$ is bounded above by\cite{BB}
\begin{equation}
\lambda_n \leq \sup_{\psi \in S_n} Q[\psi] ,
\end{equation}
where $S_n$ is any $n$-dimensional subspace of the Sobolev space for $O$. We then wish to show that $\lambda_n < b$ for some bound $b$: $O=H-E_0$ and $b=0$ for localization below the infimum of the spectrum, or $O = (H - E_g)^2$ and $b = (E_i - E_g)^2$ for localization in the gap from edge $i$. This is equivalent to proving that the Hermitian form $B[\psi,\phi] = \langle \psi, (O - b) \phi \rangle$ is negative-definite for \emph{some} $n$-dimensional subspace $S_n$ (i.e., $B[\psi,\psi] < 0$ for all $\psi \in S_n$).

If $E_i$ is $D$-fold degenerate, with degenerate generalized eigenfunctions $\psi_i^\ell$ for $\ell = 1,\ldots,D$, then the analogue of our previous approach is to form the trial functions $\psi^\ell = \gamma \psi_i^\ell$ (whose span is a subspace $S_D$), compute the $D \times D$ matrix $\mathcal{B}_{\ell \ell'} = B[\psi^\ell, \psi^{\ell'}]$, and check whether it is  negative definite as $\alpha \to 0$. We wish to find the largest subspace $S_n$ of $S_D$ for which $B$ is negative-definite, which corresponds to the number $n$ of negative eigenvalues of $\mathcal{B}$: this will be the number $n$ of localized states that are guaranteed by the theorem.

For localization below the infimum of the spectrum by a $V$ satisfying
(\ref{eq:bsbiV2}), following exactly the same steps as in
Sec.~\ref{sec:dDim}, proving that $B[\psi,\phi]$ is
negative-definite in this subspace reduces to a generalization of
condition (\ref{eq:bsbiV1}). Specifically, showing $B[\psi,\psi] < 0$
in the subspace for $\alpha \to 0$ reduces, via
(\ref{eq:final-inf-eq}), to showing that $\int V |\psi|^2 < 0$ for
every $\psi$ in the subspace.  In other words, the Hermitian form
$A[\psi,\phi] = \< \psi, V \phi \>$ must be negative-definite in
$S_n$.  In the $\alpha \to 0$ limit, this corresponds to checking the
eigenvalues of the $D\times D$ matrix $\mathcal{A}_{\ell \ell'} =
A[\psi_0^\ell, \psi_0^{\ell'}]$: the number of negative eigenvalues of
$\mathcal{A}$ is precisely the dimension of the largest
negative-definite subspace $S_n$, and hence is the number of bound
states that are guaranteed to be localized below the ground state of
$V_0$.  If we happen to have a strictly non-positive $V \le 0$, then
the Hermitian form $A$ is automatically negative-definite and we are
guaranteed $D$ localized modes.

For localization in a gap by a $V$ satisfying (\ref{eq:Vprop2}),
pulling states from band-edge $i$, following exactly the same steps as
in Sec.~\ref{sec:bssg}, one finds that $B[\psi,\phi]$ being
negative-definite reduces to a generalization of condition
(\ref{eq:Vtermsrem}): the Hermitian form $G[\psi,\phi] = \< \psi, [V^2
  + 2(E_i - E_g) V] \phi \>$ must be negative-definite in $S_n$. As
above, this simplifies for $\alpha \to 0$ to counting the number of
negative eigenvalues of the $D\times D$ matrix $\mathcal{G}_{\ell
  \ell'} = G[\psi_i^\ell, \psi_i^{\ell'}]$.  The number of negative
eigenvalues is the number of solutions that are guaranteed to be
localized from band-edge $i$. If $V$ has sign everywhere opposite to
$E_i - E_g$ and is sufficiently small (to overwhelm the $V^2$ term),
then $D$ eigenstates will be localized from this band edge.

This analysis appears to be closely related to the asymptotic
technique of \citeasnoun{We}, which also relates a number of bound
modes to the number of eigenvalues of a given sign of a small matrix,
via the Birman--Schwinger principle in the limit of weak
perturbations, but that work only explicitly considered localization
below the ground state of translation-invariant elliptic and
Schr\"odinger-type unperturbed operators.

\section{Concluding Remarks}
\label{sec:future}
Although the existence of localized solutions from defects in periodic potentials and the effective-mass analogy with the vacuum case are well known as a practical computational and experimental matter, it is gratifying to have a general, explicit proof that localization in one and two dimensions occurs in a similar manner to localization in vacuum.  A number of directions suggest themselves for future research. Although the simplicity of an elementary proof based on the min--max/variational theorem has its own appeal, the application of more sophisticated methods such as those of \citeasnoun{Si} may reveal additional information about the nature of the localized state (such as its asymptotic localization length) that cannot be gleaned from a simple variational analysis.  We would also like to transfer these results  from the Schr\"odinger (scalar) picture to the Maxwell (vector) one, in the context of localization in band gaps of photonic crystals such as photonic-crystal fibers.\cite{JoannopoulosJo08-book} A similar generalization to electromagnetism was already obtained for localization below the infimum of the spectrum (corresponding to total internal reflection in Maxwell's equations) for non-periodic\cite{BB} and periodic\cite{KAJ} media, and in photonic band gaps for sufficiently large defects.\cite{KO,Miao08}

For localization in gaps, we should remark that the condition (\ref{eq:Vprop3}) on the size of the perturbation $V$ is somewhat unsatisfying.  Intuitively, a ``small'' perturbation $V$ could be one where either $|V|$ is small at every point or where $|V|$ is not small but the support of $V$ is small. The latter case, however, of a large $|V|$ with a small integral, violates our smallness condition (\ref{eq:Vprop3}) for a sufficiently large $|V|$ no matter how small the support might be. This does not mean that there are no localized states in that limit---our proof only gives a sufficient condition for localization, not a necessary condition---but it suggests that some reformulation to handle this physically interesting possibility might be desirable.

\begin{acknowledgments}
\label{sec:ack}
We are grateful to Fritz Gesztesy, Dirk Hundertmark, Patrick Lee,
Leonid Levitov, Emil Prodan, Alexander Pushnitski, Barry Simon, and
Michael Weinstein for helpful discussions and references to the
literature. This work was supported in part by the Macaulay Honors
College at the City University of New York.
\end{acknowledgments}

\section*{Appendix}
\label{sec:appendix}

Here we will provide more details for the calculations of Secs.~\ref{sec:bsbi} and~\ref{sec:bssg}. Let $\psi$ be an eigenstate multiplied by a \emph{real} function $\gamma$ whose derivatives decay quickly enough, i.e. $\psi = \psi_0 \gamma$ where and $\gamma$ is real (we can think of the example used in this paper) and $\psi_0$ is bounded and satisfies $H_0 \psi_0 = E_0 \psi_0.$ We will explore some of the terms in both $\< \psi, (H - E_0) \psi \>$ and $\< (H-E_g) \psi, (H-E_g) \psi \>.$ Our goals in the above sections were to show that most terms tended to zero. In order for this to be apparent, we had to perform several algebraic manipulations of the integrals that are given here more explicitly. 

\subsection{Calculations for Sec.~\ref{sec:bsbi}}
\label{sub:bsbiappendix}

We will derive the general formula used in Sec.~\ref{sec:bsbi} first. We have 
\begin{align}
(H-E_0) \psi &= (-\nabla^2 \psi_0 + V_0 \psi_0) \gamma - E_0 \psi_0 \gamma \nonumber \\
        &\qquad + V \psi - 2 \nabla \psi_0 \cdot \nabla \gamma - \psi_0 \nabla^2 \gamma \nonumber \\
	&= V \psi - 2 \nabla \psi_0 \cdot \nabla \gamma - \psi_0 \nabla^2 \gamma
\end{align}
since $(-\nabla^2 + V_0) \psi_0 = E_0 \psi_0,$
so that $U[\psi] = \< \psi, (H-E_0)\psi \>$ is
\be
U[\psi] = \int V |\psi|^2 - 2 \int \gamma \psi_0^* \nabla \psi_0 \cdot \nabla \gamma - \int |\psi_0|^2 \gamma \nabla^2 \gamma.
\label{eq:bsbigen1}
\end{equation}
Because $U[\psi]$ must always be real and $\gamma$ is assumed to be real, the imaginary part of $\psi_0^* \nabla \psi_0$ must integrate to zero. Therefore, we can replace $2 \psi_0^* \nabla \psi_0$ by $\psi_0^* \nabla \psi_0 + \psi_0 \nabla \psi_0^*$ in the integrand. We can also use the identity $\gamma \nabla \gamma = \frac{1}{2} \nabla (\gamma^2)$ so that equation (\ref{eq:bsbigen1}) becomes
\be
\int V |\psi|^2 -  \frac{1}{2} \int \nabla |\psi_0|^2 \cdot \nabla (\gamma^2) - \int |\psi_0|^2 \gamma \nabla^2 \gamma. 
\label{eq:bsbigen2}
\end{equation}
We now rewrite middle term in another manner in order to eliminate the slowly decaying second derivative of $\gamma$, using integration by parts:
\begin{align}
\int \nabla |\psi_0|^2 \cdot \nabla (\gamma^2) &= \int_{\p} |\psi_0|^2 \nabla (\gamma^2) - \int |\psi_0|^2 \nabla^2 (\gamma^2) \\
	&=  -\int |\psi_0|^2 \nabla^2 (\gamma^2),
\end{align}
where ``$\int_{\p}$'' stands for the boundary integral.
The boundary integral is zero because the $\gamma$ term and it first derivative decay fast enough and $\psi_0$ is bounded. 
Therefore, all we have to show in Sec.~\ref{sec:bsbi} is that
\be
\int V |\psi|^2 + \frac{1}{2} \int |\psi_0|^2 \nabla^2 (\gamma^2) - \int |\psi_0|^2 \gamma \nabla^2 \gamma < 0 
\ee
for some sufficiently small choice of the parameter $\alpha$.

\subsection{Calculations for Sec.~\ref{sec:bssg}}
\label{sub:bssgappendix}

We will now derive the general formula used in Sec.~\ref{sec:bssg}. The same assumptions hold as in the previous subsection, but with minor modifications from Sec.~\ref{sec:bssg}, for example $\psi = \psi_i \gamma$ where $i=1,2$ signifies the lower or upper edge of the band gap, respectively. First, we calculate $(H-E_g) \psi.$ 
\be
(H-E_g) \psi = (\Delta E) \psi + V \psi - 2 \nabla \psi_i \cdot \nabla \gamma - \psi_i \nabla^2 \gamma,
\end{equation}
where $\Delta E = E_i - E_g$ as in Sec.~\ref{sec:bssg}.
Then the generalized equation for $\lVert (H-E_g)\psi \rVert^2$ is given by (employing some of the trivial simplifications from the previous section)
\begin{align}
& (\Delta E)^2 \lVert \psi \rVert^2 + 2(\Delta E) \int V |\psi|^2 + \lVert V \psi \rVert^2 \label{eq:Topgen1} \\
&\qquad - \int V \nabla |\psi_i|^2 \cdot \nabla (\gamma^2) 
  - 2 \int V |\psi_i|^2 \gamma \nabla^2 \gamma  \label{eq:Topgen2}\\
&\qquad -\; 2 (\Delta E) \int  |\psi_i|^2 \gamma \nabla^2 \gamma \label{eq:Middlegen1}\\
&\qquad +\; \lVert \psi_i \nabla^2 \gamma \rVert^2 +  2 \int \nabla^2 \gamma \nabla |\psi_i|^2 \cdot \nabla \gamma  \label{eq:Middlegen2} \\
&\qquad -\; (\Delta E) \int \nabla |\psi_i|^2 \cdot \nabla (\gamma^2) + 4 \lVert \nabla \psi_i \cdot \nabla \gamma \rVert^2. \label{eq:Bottomgen} 
\end{align}
Notice that the first term in (\ref{eq:Bottomgen}) is similar to the middle term in~(\ref{eq:bsbigen2}) from the previous subsection, so the same analysis shows that (\ref{eq:Bottomgen}) becomes
\be
+\;(\Delta E)  \int |\psi_i|^2 \nabla^2 (\gamma^2) + 4 \lVert \nabla \psi_i \cdot \nabla \gamma \rVert^2.
\label{eq:Bottomgen2}
\end{equation}
In (\ref{eq:Top}) of Sec.~\ref{sec:bssg}, this first term of
(\ref{eq:Bottomgen2}) is combined with
term~(\ref{eq:Middlegen1}). Several of the other terms are rearranged
to make the presentation more concise.

\bibliographystyle{ieeetr}
\bibliography{localization}

\end{document}